\documentclass[conference,review]{IEEEtran}

\IEEEoverridecommandlockouts

\usepackage{cite}
\usepackage{amsmath,amsfonts,amsthm}
\usepackage{algorithm,algorithmic}
\usepackage{xspace}
\usepackage{graphicx}
\usepackage{textcomp}
\usepackage{xcolor}
\usepackage{listings}
\lstdefinestyle{Python}
{
    basicstyle=\footnotesize\ttfamily,
    numberblanklines=false,
    language=python,
    tabsize=2,
    commentstyle=\color{gray},
    keywordstyle=\bfseries\color{eclipsePurple},
    morekeywords={assert},
    stringstyle=\color{eclipseBlue},
    columns=flexible,
    identifierstyle=,
}
\definecolor{eclipseBlue}{RGB}{42,0.0,255}
\definecolor{eclipseGreen}{RGB}{63,127,95}
\definecolor{eclipsePurple}{RGB}{127,0,85}
\usepackage{multirow}
\usepackage{pifont}
\usepackage{hyperref}
\newcommand{\mytitle}{Assessing Reliability of Statistical Maximum Coverage Estimators in Fuzzing}
\hypersetup{pdftitle={\mytitle},
colorlinks=true,linkcolor=blue,citecolor=blue,filecolor=blue,urlcolor=blue
}
\usepackage{tikz}
\usepackage{pgfplots}
\usepackage{balance}
\pgfplotsset{compat=1.9}
\usetikzlibrary{fit}
\usetikzlibrary{positioning}
\usetikzlibrary{datavisualization,datavisualization.formats.functions}
\usetikzlibrary{shapes, calc, decorations, automata}
\usetikzlibrary{shapes.geometric, arrows}
\usetikzlibrary{decorations.markings}
\usetikzlibrary{arrows.meta}
\usetikzlibrary{chains,shadows.blur}
\usetikzlibrary{math}
\usepgfplotslibrary{fillbetween}
\tikzstyle{startstop} = [rectangle, rounded corners, minimum width=3cm, minimum height=1cm,text centered, draw=black, fill=red!30]
\tikzstyle{process} = [rectangle, rounded corners, minimum width=3cm, minimum height=1cm, text centered, draw=black, fill=orange!30, inner sep=5pt,minimum size=10pt]
\tikzstyle{decision} = [diamond, minimum width=3cm, minimum height=1cm, text centered, draw=black, fill=green!30]
\tikzstyle{arrow} = [thick,->,>=stealth]
\usepackage{cleveref}
\usepackage{subcaption}

\usepackage{pifont}
\usepackage{enumitem}
\usepackage{url}
\usepackage{booktabs}
\usepackage{tabularx}
\usepackage{multirow,array}
\usepackage{threeparttable}
\usepackage{bm}
\usepackage{syntax}
\def\|#1|{\textit{#1}}
\def\<#1>{\texttt{#1}}
\def\[[#1\]]{\texttt{#1}}

\def\term#1{\texttt{'\textbf{#1}'}}
\def\nonterm#1{\textlangle\textnormal{\emph{#1}}\textrangle}

\def\expandsto{\(\rightarrow{}\)}
\usepackage{bbding}
\DeclareUnicodeCharacter{1F3B2}{\dice}

\usepackage{wrapfig}
\usepackage{tikz}
\usetikzlibrary{shapes, calc, decorations, automata}
\usetikzlibrary{shapes.geometric, arrows}
\usetikzlibrary{decorations.markings}
\usetikzlibrary{arrows.meta}
\usetikzlibrary{chains,shadows.blur}

\newcounter{todocounter}

\newcommand{\done}[1]{\marginpar{$*$}\textcolor{green}{\stepcounter{todocounter}\footnote[\thetodocounter]{\textcolor{black}{\textbf{DONE }}\textit{#1}}}}
\newcommand{\rebuttal}[1]{#1}

\renewcommand{\done}[1]{} 

\usepackage{tikz}

\def\BibTeX{{\rm B\kern-.05em{\sc i\kern-.025em b}\kern-.08em
    T\kern-.1667em\lower.7ex\hbox{E}\kern-.125emX}}

\newtheorem*{theorem*}{Theorem}

\usepackage{tcolorbox}
\definecolor{mycolor}{rgb}{0.122, 0.435, 0.698}
\definecolor{gray1}{gray}{0.3}

\definecolor{codegreen}{rgb}{0,0.6,0}
\definecolor{codegray}{rgb}{0.5,0.5,0.5}
\definecolor{codepurple}{rgb}{0.58,0,0.82}
\definecolor{backcolour}{rgb}{0.95,0.95,0.92}
\lstdefinestyle{mystyle}{
    commentstyle=\color{codegreen},
    keywordstyle=\color{magenta},
    numberstyle=\tiny\color{codegray},
    stringstyle=\color{codepurple},
    basicstyle=\tiny\ttfamily,
    breakatwhitespace=false,
    breaklines=true,
    captionpos=b,
    keepspaces=true,
    numbers=left,
    numbersep=5pt,
    showspaces=false,
    showstringspaces=false,
    showtabs=false,
    tabsize=2,
    columns=fixed
}
\definecolor{darkgreen}{rgb}{0.0, 0.5, 0.0}
\definecolor{darkred}{rgb}{0.82, 0.1, 0.26}

\begin{document}

\title{\mytitle}

\author{
\IEEEauthorblockN{Danushka Liyanage\IEEEauthorrefmark{1}, Nelum Attanayake\IEEEauthorrefmark{1}, Zijian Luo\IEEEauthorrefmark{1}, Rahul Gopinath}
\IEEEauthorblockA{%
\textit{School of Computer Science, University of Sydney, Australia} 
\thanks{\textsuperscript{*}These authors contributed equally and share first authorship.}
}}


\maketitle

\thispagestyle{plain}
\pagestyle{plain} 


\begin{abstract}
\textbf{Background:} Fuzzers are often guided by coverage, making the estimation of maximum 
achievable coverage a key concern in fuzzing.
However, achieving 100\% coverage is infeasible for most real-world
software systems, regardless of effort.
While static reachability analysis can provide an upper bound,
it is often highly inaccurate.
Recently, statistical estimation methods based on species richness
estimators from biostatistics have been proposed as a potential solution.
Yet, the lack of reliable benchmarks with labeled ground truth has
limited rigorous evaluation of their accuracy.






\textbf{Objective:}
This work examines the reliability of reachability estimators from two axes:
addressing the lack of labeled ground truth and evaluating their reliability on real-world programs.

\textbf{Methods:}
  (1) To address the challenge of labeled ground truth, we propose an evaluation
framework that synthetically generates large programs with complex
control flows, ensuring well-defined reachability and providing ground truth for
evaluation.
  (2) To address the criticism from use of synthetic benchmarks, we adapt a
reliability check for reachability estimators on real-world benchmarks without
labeled ground truth---by varying the size
of sampling units, which, in theory, should not affect the estimate.

\textbf{Results:}
These two studies together will help answer the question of whether current
reachability estimators are reliable, and defines a protocol to evaluate future
improvements in reachability estimation.
\end{abstract}

\section{Introduction}

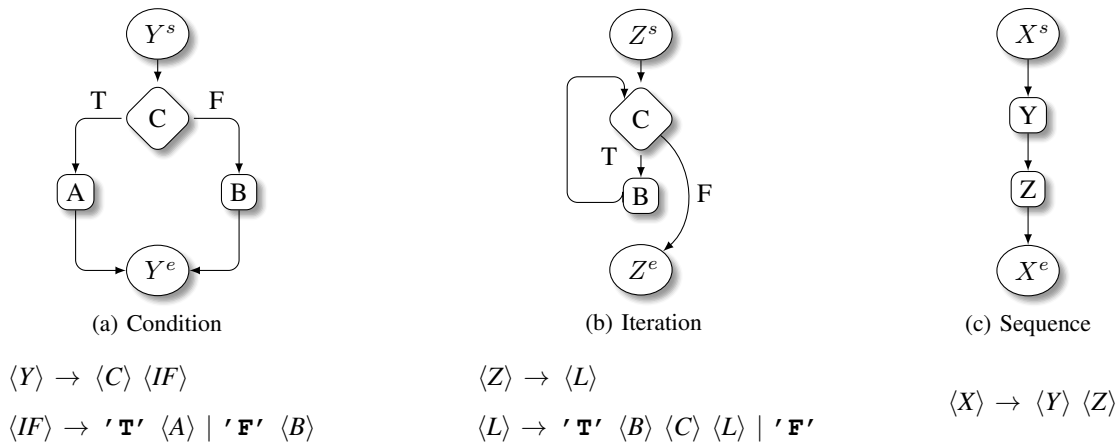
\begin{figure*}
\centering
\begin{subfigure}[h]{0.35\textwidth}
\centering
\begin{tikzpicture}[auto,
  node distance = 3mm,
  start chain = going below,
  start/.style = {ellipse,draw,text width= 1em,rounded corners,blur shadow,fill=white,
        on chain,align=center},
  box/.style = {draw,rounded corners,blur shadow,fill=white,
        on chain,align=center},
  cond/.style = {diamond,draw,rounded corners,blur shadow,fill=white,
        on chain,align=center},
  stop/.style = {ellipse,draw,text width= 1em,rounded corners,blur shadow,fill=white, on chain,align=center},
]
 \node[start] (b1)    {$Y^s$};
 \node[cond] (b2)    {C};
 \node[box, below left=0.5cm and 0.6cm of b2] (b21) {A};
 \node[box, below right=0.5cm and 0.6cm of b2] (b22) {B};
 \node[stop,below=12mm of b2] (b3)    {$Y^e$};
 \begin{scope}[rounded corners,-latex]
 \path
 (b1) edge (b2);
 \draw[-latex] (b2) -| node[pos=0.25,above] {T} (b21);
 \draw[-latex] (b2) -| node[pos=0.25,above] {F} (b22);
  \draw[-latex] (b21) |- (b3);
  \draw[-latex] (b22) |- (b3);
\end{scope}
\end{tikzpicture}
\caption{Condition}
\label{tikz:condition}
\end{subfigure}
\begin{subfigure}[h]{0.35\textwidth}
\centering
\begin{tikzpicture}[auto,
  node distance = 3mm,
  start chain = going below,
  start/.style = {ellipse,draw,text width= 1em,rounded corners,blur shadow,fill=white, on chain,align=center},
  box/.style = {draw,rounded corners,blur shadow,fill=white,
  on chain,align=center},
  cond/.style = {diamond,aspect=1,draw,rounded corners,blur shadow,fill=white,
  on chain,align=center},
  stop/.style = {ellipse,draw,text width= 1em,rounded corners,blur shadow,fill=white, on chain,align=center},
  ]
 \node[start] (b1)    {$Z^s$};      
 \node[cond] (b2)    {C};      
 \node[box] (b3)    {B};  
 \node[stop,below=4mm of b3] (b4)    {$Z^e$};
 \begin{scope}[rounded corners,-latex]
  \path (b2.-40) edge[bend left=50] node[midway,right] {F} (b4.40)
  (b1) edge node [midway,below, xshift=-12pt,yshift=-25pt] {T}(b2)
  (b2) edge (b3);
  \draw (b3.200) -- ++(0,0) -| ([xshift=-5mm]b2.west) |-
  ([yshift=3mm]b2.130) -- (b2.130);
 \end{scope}
\end{tikzpicture}
\caption{Iteration}
\label{tikz:iteration}
\end{subfigure}
\begin{subfigure}[h]{0.2\textwidth}
\centering
\begin{tikzpicture}[auto,
  node distance = 3mm,
  start chain = going below,
  start/.style = {ellipse,draw,text width= 1em,rounded corners,blur shadow,fill=white,
        on chain,align=center},
  box/.style = {draw,rounded corners,blur shadow,fill=white,
        on chain,align=center},
  stop/.style = {ellipse,draw,text width= 1em,rounded corners,blur shadow,fill=white, on chain,align=center},
  ]
 \node[start] (b1)    {$X^{s}$};
 \node[box,below=5mm of b1] (b2)    {Y};
 \node[box,below=5mm of b2] (b3)    {Z};
 \node[stop,below=5mm of b3] (b4)    {$X^{e}$};
 \begin{scope}[rounded corners,-latex]
  \path
  (b1) edge (b2)
  (b2) edge (b3)
  (b3) edge (b4);
 \end{scope}
\end{tikzpicture}
\caption{Sequence}
\label{tikz:sequence}
\end{subfigure}
\vspace{10pt}

\begin{subfigure}[h]{0.35\textwidth}
\begin{grammar}\centering
    <Y> $\rightarrow$ <C> <IF> \phantom{aaaaaaaaaa}
    
    <IF> $\rightarrow$ \term{T} <A> | \term{F} <B>
\end{grammar}
\end{subfigure}
\begin{subfigure}[h]{0.35\textwidth}
\begin{grammar}\centering
  <Z> $\rightarrow$ <L> \phantom{aaaaaaaaaaaaaaaaaa}
  
  <L> $\rightarrow$ \term{T} <B> <C> <L> | \term{F}
\end{grammar}
\end{subfigure}
\begin{subfigure}[h]{0.2\textwidth}
\begin{grammar}\centering
  <X> $\rightarrow$ <Y> <Z>
\end{grammar}
\end{subfigure}
\caption{The basic control-flow structures}
\label{fig:cfg}
\end{figure*}

Fuzzing is an automated testing technique that leverages guidance provided
by coverage~\cite{boehme2016coverage}.
The budget allocated to fuzzing, and hence, the stopping criterion,
is determined by the maximum achievable coverage~\cite{fell2017review}.
However, achieving 100\% coverage in real-world programs is
nearly impossible~\cite{fraser2013whole,horgan1994achieving,boehme2018stads}.

Yet, quantifying the maximum reachable coverage (also called
\emph{maximum reachability}) remains a challenge,
which raises the fundamental question:
\emph{How close is a fuzzing campaign to its maximum reachable coverage?}

Researchers have recently adapted \emph{species richness estimators} from
biostatistics~\cite{chao2016species} as a principled approach to estimate
the reachable coverage of a fuzzing campaign~\cite{boehme2018stads}.
These estimators model fuzzing as a statistical sampling process where
each test input exercises one or more program behaviors, such as reaching
a program element. This process can be used to estimate the total number of
such behaviors~\cite{boehme2018stads}.

However, evaluating the accuracy and reliability of such estimators is
difficult because evaluating such estimators requires benchmarks
composed of several large, complex programs where maximum
reachability is known \cite{liyanage2021security}.
Unfortunately, establishing the maximum reachability is impossible except
for trivial programs.
Hence, researchers have resorted to using small programs, fuzzed to saturation,
as an alternative~\cite{liyanage2023reachable}.
The limitation here is that the performance of estimators in small
programs may not generalize to complex real-world software.

This paper suggests an alternative approach.
We observe that 
a program's control flow, along with its data flow, determines its overall complexity,
and hence, the difficulty in achieving coverage.
Moreover, the control flow of any structured program can be represented
as a context-free LL(1)\footnote{
An LL(1) CFG is a CFG supporting one look-ahead token.}
grammar with labeled edges~(see for e.g., \Cref{fig:cfg}).
We further observe that the reverse is also true: converting an LL(1)
grammar to a recursive descent parser is a well-known, automated process.
The conversion produces parsers with control-flow structures comparable to
the original program. In other words, the space of LL(1) grammars
can be viewed as equivalent to the space of possible program control-flows.

It is possible to generate arbitrary LL(1) grammars.
Moreover, we can fine-tune the complexity of such generated grammars
based on several dimensions such as the size of the grammar,
the number of direct and indirect recursions allowed, the
number of linear-recursive rules, among other parameters.

We propose to leverage these synthetically generated, complex,
context-free grammars by converting them into parsers,
and use these as benchmarks for
verifying reliability of reachability estimators.
Hence, this is a \emph{exploratory study} asking:

\noindent\textbf{RQ1:} How accurate are the maximum reachability estimators?

Our hypothesis is that when using these generated programs,
the examined estimators will consistently generate maximum
reachability estimates such that the true reachability value
will lie within the 90\% confidence-interval (CI) of the estimate.


The synthetic benchmarks can be critiqued that they are
are different from real-world programs.
Hence, we wonder, can complex real-world programs still be used for evaluation,
even though their true maximum reachability is unknown?

A recent research on evaluating the reliability of biostatistics based
estimators for mutation analysis~\cite{Kuznetsov2024empirical}
provides a solution.
The researchers argue that the choice of \emph{sampling units} in species
richness estimators \emph{should not} significantly affect the final
point estimate. The authors point out that changing the granularity of
sampling units thus provide a mechanism to verify the reliability of
a species richness estimator.

In prior research on coverage estimation~\cite{liyanage2023reachable},
the sampling unit was a fixed time interval (e.g., 15 minutes).
Since this choice is arbitrary, we can change the time interval to
larger units, for example 60 minutes. Changing the interval's granularity
should, in theory, not impact the final estimate. We expect that
when the sampling unit granularity is changed, the CIs
of each estimates significantly overlap, and the point estimate from one
is within the CI of the other.

Hence, this is a second \emph{exploratory study} asking:

\noindent\textbf{RQ2:} Are the maximum reachability estimators for $S$
reliable when sampling unit size $r$ is changed?

Our hypothesis is that the point estimates from different sampling unit
granularities will consistently lie within the 90\% CI 
of each other. \emph{If not}, then it is likely that
the estimators are unreliable, and provide a secondary evaluation of the
accuracy of these estimators using real world subjects.


\noindent{}\textbf{Contributions:}

\begin{itemize}
  \item \textbf{Synthetic benchmarks for evaluation:} We propose a novel
framework for generating synthetic program benchmarks that emulate real-world
programs, and use this to evaluate maximum reachability estimators.

  \item \textbf{Sampling-unit based evaluation of reliability:} We will evaluate
    the reliability of maximum reachability estimators on 32 real-world programs
    by comparing the estimates under sampling units of diverse granularity.

\end{itemize}

The remainder of this paper is organized as follows. \Cref{sec:preliminaries} introduces the statistical reachability estimation in fuzzing. \Cref{sec:method} presents our methodology for assessing the accuracy and reliability of statistical estimators of maximum reachability. \Cref{sec:setup} details the experimental setup used to evaluate our approach. Then, \Cref{sec:related} and \Cref{sec:threats} discuss the associated related work and threats to validity. \Cref{sec:future} discusses future work, and
\Cref{sec:conclusion} concludes.

\section{Preliminaries} \label{sec:preliminaries}
To evaluate fuzzer effectiveness, we need the maximum reachability in a given program, which
often need to be approximated. 
Major approximation approaches include:

\noindent\textbf{Non-statistical approaches:}
\begin{enumerate}
    \item \emph{Plateaued Coverage~\cite{boehme2018stads}:}
      Previous research treated the observed coverage at the end of a fuzzing
      campaign when coverage plateaued as the ground truth.
      However, given the exponential difficulty of finding new
      behaviors~\cite{boehme2021residual},
      this approach is unsatisfactory.
    \item \emph{Small-scale Programs~\cite{liyanage2023reachable}:}
      For programs with limited behavioral diversity,
      fuzzers often reach coverage saturation quickly.
      The remaining program elements may be inspected
      manually to determine the reachability.
      However, they lack the complexity of real-world programs,
      and are a poor substitute for good benchmarks.
\end{enumerate}

\noindent\textbf{Statistical approaches:}
\begin{enumerate}
    \item \emph{Bootstrapping~\cite{liyanage2023reachable}:}
    When fuzzing large real-world programs, once coverage stabilizes,
    bootstrapping/resampling from campaign data can statistically
    approximate ground truth.
    However, it relies on assumptions such as independence of inputs,
    and relative homogeneity of behavioral complexity in unexplored regions.
    \item\emph{Species richness estimators:}
    If we can partition the input space into sampling units, we can
    keep track of distinct program elements reached in each unit. From this,
    we can estimate the number of program elements that are yet to be seen in the limit~\cite{boehme2018stads}.
    The idea is to leverage the species diversity estimators from
    biostatistics which are used to estimate unseen species from a survey
    of known species~\cite{chao2017thirty}.
    While promising, species-richness estimators have similar assumptions as
    that of bootstrapping, which may not be reasonable in real-world programs.
\end{enumerate}
%
We next discuss the theoretical statistical approach in detail.
\subsection{Probabilistic Model for Fuzzing}
The STADS\footnote{STADS—Software Testing As Discovery of Species} framework~\cite{boehme2018stads,boehme2021residual,nguyen2022bedivfuzz,boehme2020boosting} formalizes
fuzzing as a statistical sampling process $\mathcal{F}$. Each test input is drawn with replacement from the program's input space $\pmb{\mathcal{D}}$. Assuming a sequence of $N$ independent and identically distributed (i.i.d.) random variables are drawn, we formally define the fuzzing campaign $\mathcal{F}$ as:

\begin{equation*}
    \mathcal{F}=\{X_n \mid X_n \in \pmb{\mathcal{D}}\}_{n=1}^N
\end{equation*}

The input space $\pmb{\mathcal{D}}$ consists of $S$ (potentially overlapping) subdomains $\{\mathcal{D}_i\}_{i=1}^S$, each representing a distinct \emph{coverage element}. An input $X_n \in \mathcal{F}$ is said to discover a \emph{new} coverage element $\mathcal{D}_i$ if $X_n \in \mathcal{D}_i$ and no earlier input $X_m \in \mathcal{F}$ for $m < n$ has been drawn from $\mathcal{D}_i$—that is, $\mathcal{D}_i$ is being encountered for the first time.

\begin{figure*}[ht] 
  \centering
\begin{subfigure}[h]{0.45\textwidth} 
  \centering
\lstset{numbers=left,xleftmargin=2em, numberstyle=\color{lightgray}
} 
\begin{lstlisting}[style=Python, escapechar=|,numbersep=2pt]
def bsearch(x, v, n):
  low, high = 0, n-1
  while low <= high:
    mid=(low+high)/2
    if x < v[mid]:
      high=mid-1
    else:
      if x > v[mid]:
        low=mid+1
      else:
        return mid
  return None
\end{lstlisting}
\caption{\<bsearch> program}
\label{fig:bsearch1}
\end{subfigure}
\begin{subfigure}[h]{0.45\textwidth}   
  \centering
\begin{grammar}
  <bsearch> $\rightarrow$ \term{l.2} <while.3> \term{l.12}

  <while.3> $\rightarrow$ \term{l.3}
   \alt \term{l.3} \term{l.4} \term{l.5} <if.5> <while.3>

  <if.5>  $\rightarrow$  \term{T} \term{l.6}
   \alt \term{F} \term{l.8} <if.8>

  <if.8> $\rightarrow$ \term{T} \term{l.9}
   \alt \term{F} \term{l.11}
\end{grammar} 
\caption{\<bsearch> control-flow equivalent grammar}
\label{fig:bsearch3}
\end{subfigure}
  \caption{Equivalent LL(1) grammar for \<bsearch> control-flow. Tokens such as \term{l.3} represent statements in the original code.}
\label{fig:bsearch}
\end{figure*}

\subsection{Bernoulli Product Model}
Since each input in $\mathcal{F}$ may belong to one or more coverage elements, the STADS model represents coverage using \emph{sampling-unit-based incidence data}~\cite{colwell2012models,chao2017thirty}, where a sampling unit aggregates all inputs generated within a fixed time interval. The underlying probabilistic model for this representation is the \emph{Bernoulli product model}. Grouping inputs into sampling units is essential to reduce the overhead of tracking fine-grained coverage information for each individual input during fuzzing.

For each sampling unit, the collected data indicates whether a coverage element
has been reached. Let $\pi_i$ denote the probability that a sampling unit covers
element $\mathcal{D}_i$, assuming $\pi_i$ remains constant across all randomly
selected sampling units.
In general, the sum of all $\pi_i$ values does \emph{not} equal unity.

During a fuzzing campaign, suppose we record $t$ sampling units. The incidence data forms an element$\times$sampling-unit incidence matrix ${W_{ij};i=1,2,\dots,S,j=1,2,\dots,t}$ with $S$ rows and $t$ columns, where $W_{ij} = 1$ if element $i$ is covered in sampling unit $j$, and $W_{ij} = 0$ otherwise.

The incidence frequency $Y_i$ represents the number of sampling units in which element $\mathcal{D}_i$ is covered; i.e., $Y_i=\sum_{j=1}^{t}W_{ij}$. A coverage element $\mathcal{D}_i$ that has not been covered by any sampling unit will have an incidence frequency of zero; i.e., $Y_i=0$.

Given the set of detection probabilities $(\pi_1,\pi_2,\dots,\pi_S)$, we assume each element $W_{ij}$ in the incidence matrix follows a Bernoulli distribution with probability $\pi_i$. The probability distribution for the incidence matrix is:

\begin{equation}
    \begin{split}
        P(W_{ij}=w_{ij};i=1,2,\dots,S,j=1,2,\dots,t) \\
        = \prod_{j=1}^{t}\prod_{i=1}^{S}\pi_i^{w_{ij}}(1-\pi_i)^{1-w_{ij}} \\
        = \prod_{i=1}^{S}\pi_i^{y_i}(1-\pi_i)^{t-y_i}.
    \end{split}
\end{equation}

The marginal distribution for the incidence-based frequency $Y_i$ for the $i$-th coverage element follows a binomial distribution characterized by $t$ and the detection probability $\pi_i$:

\begin{equation}
    P(Y_i=y_i) = \binom{t}{y_i}\pi_i^{y_i}(1-\pi_i)^{t-y_i}, \qquad i=1,2,\dots,S.
\end{equation}

Denote the incidence frequency counts by $(f_0, f_1, \dots, f_t)$, where $f_k$ is the number of elements covered in exactly $k$ sampling units in the data, $k=0,1,\dots,t$. Here, $f_1$ represents the number of \emph{singleton} elements (those that are covered in only one sampling unit), and $f_2$ represents the number of \emph{doubleton} elements (those that are covered in exactly two sampling units). The unobservable zero frequency count $f_0$ denotes the number of coverage elements that are not covered by any of the $t$ sampling units. Then, the number of covered elements in the current campaign is $S(t)=\sum_{i>0}f_i$, and $S(t)+f_0=S$.

\section{Subjects} \label{sec:method}
This study is based on two different benchmarks, one synthetic---but with known ground truth, and
another a real-world benchmark---but with unknown ground truth.
\subsection{Synthetic Benchmarks With Known Reachability}
We generate large programs with complex control flow by leveraging the
structural isomorphism between control-flow graphs and context-free grammars.

\noindent\textbf{Primer on Context-Free Grammars.}
A context-free grammar (also called a \emph{grammar})
is a formal system for defining a language. The following is a simple grammar:
\begin{grammar}\centering
  <E> $\rightarrow$ <D> <Es>\phantom{a}\phantom{a}\phantom{a}\phantom{a}\phantom{a}\phantom{a}\phantom{a}\phantom{a}

<Es> $\rightarrow$ \term{+} <D> <Es> $\mid$ $\epsilon$

<D> $\rightarrow$ \term{0} $\mid$ \term{1} $\mid \ldots$ 
\phantom{a}
\end{grammar}
%

The \emph{terminal symbols} are the vocabulary of the language.
In the above, \term{0} is a terminal symbol.
The grammar also contains \emph{nonterminal symbols} which are named
constructs that can be expanded into other terminal or nonterminal
symbols (e.g., \nonterm{E} in the above).

Nonterminal symbols are expanded into other symbols according to a
set of \emph{production rules} (also called \emph{rules}),
which together \emph{define} the nonterminal.
The definition of the nonterminal \nonterm{Es} in the above example is given
by
\mbox{\nonterm{Es} \expandsto \term{+} \nonterm{D} \nonterm{Es} $\mid$ $\epsilon$}
which consists of two rules for expanding \nonterm{Es}.

An LL(1) grammar is a special case of a context-free grammar where the production rules
in a definition can be chosen based on the next symbol to be parsed. For example, in the
definition of \nonterm{Es} the next rule to be chosen can be based on whether the
next symbol to be parsed is \term{+} or if there are no more symbols to parse ($\epsilon$).
We refer to LL(1) context-free grammars simply as \emph{grammars} from here on.

\noindent\textbf{From Control-Flow to Grammar.}
A control-flow graph represents the possible execution paths of a program and is
typically modeled as a directed graph.
\Cref{fig:cfg} illustrates example control-flow graphs for three procedures:
\<Y>, \<Z>, and \<X>.
Each graph consists of a set of nodes $N$, where each node $N_i$ represents a
program statement.
While traditionally a node corresponds to a basic block, in this work we treat each node as representing a single statement, without loss of generality.
Edges $E_{i,j}$ define possible control transfers. These are:
\begin{itemize}
    \item \textbf{Sequential}, representing sequential execution;
    \item \textbf{Conditional}, for branching (e.g., \<if-then-else>);
    \item \textbf{Loop}, modeling iteration (e.g., \<while> loops);
    \item \textbf{Call}, connecting to other procedures.
\end{itemize}

These control-flow structures map naturally to production rules in a
context-free grammar. Specifically:
\begin{itemize}
    \item Conditionals such as \<if C: A else: B> map to
      \mbox{\nonterm{$\bullet$} \expandsto \nonterm{C} \nonterm{$\text{IF}^i$}}\\
      \mbox{\nonterm{$\text{IF}^{i}$}\expandsto \term{T} \nonterm{A}  $|$ \term{F} \nonterm{B}};
    \item Loops such as \<while C: B> map to\\
      \mbox{\nonterm{$\bullet$} \expandsto \nonterm{C} \nonterm{$L^i$}}\\
      \mbox{\nonterm{$L^{i}$} \expandsto \term{T} \nonterm{B} \nonterm{C} \nonterm{$L^i$} $|$ \term{F}};
    \item Procedure definitions translate to nonterminals.
\end{itemize}
\Cref{fig:bsearch} shows the equivalent grammar from a given
program.
That is, the space of all possible context-free grammars covers the space
of all structured program control-flows.

\noindent\textbf{From Grammar to Executable Programs.}
Given any such LL(1) grammar,
we can generate a recursive-descent parser using these rules:
\begin{itemize}
  \item Each nonterminal becomes a procedure. For example, \nonterm{E} is translated to \<def parse\_E>.
  \item Each production rule is implemented in that procedure.
  \item Lookahead tokens (the next unprocessed input symbols)
  guide which rule to apply.
  \item Linear recursion can be replaced with a \<while> loop.
\end{itemize}

By following this process, we obtain a parser whose control-flow directly mirrors the structure of the grammar.
For example, given the grammar for for \term{E} we discussed previously,
%
we can translate this to:
\begin{lstlisting}[style=Python, escapechar=|,numbersep=2pt]
def parse_E():
  parse_D()

  while lookahead() == '+':
    consume('+')
    parse_D()
  return node

def parse_D():
  token = lookahead()
  if token and isdigit(token):
    consume(token)
  else:
    raise Error()
\end{lstlisting}
This parser exhibits control-flow similar to that of the program from which the grammar may have been derived.

%
%
%
%
%
%

\noindent\textbf{Controlling the program complexity.}
While generating grammars, we have several mechanisms to
precisely control the complexity. These include:

\begin{itemize}
    \item The number of nonterminals (corresponding to the number of procedures),
    \item The number and length of production rules (corresponding to branching complexity),
    \item The depth and type of recursion (direct, indirect, or linear recursion),
    \item The inclusion of unreachable nonterminals or dead code to simulate partially unreachable programs.
\end{itemize}

These parameters allow us to generate large structured parser programs with
complex control-flows. We can additionally increase the difficulty level by
adding extra constraints to the generated parser such
as context-sensitivity.
\rebuttal{We will verify that the programs thus generated are similar in \emph{cyclomatic complexity} to programs from our real-world benchmarks. We will
also report the hyper parameters used.}

To prevent trivial estimation (e.g., estimators that always return 100\%),
we introduce unreachable elements by including nonterminals that 
unreachable from the start symbol.
We may also wrap parser call-sites with guards or inject dead branches.

\subsection{Real-world Benchmarks With Unknown Reachability}
To overcome any limitation that may exist due to the synthetic nature of our
previously proposed benchmark, we propose a second criterion that can be applied
to real-world benchmarks without relying on the availability of ground truth.

Our approach is inspired by the previous research on equivalent mutant
estimation~\cite{Kuznetsov2024empirical}, which suggests that varying
the granularity of sample-units can be leveraged to evaluate the reliability of estimators.
The idea is that, while incidence-based statistical estimators of
fuzzing effectiveness may be sensitive to the sampling unit size ($r$),
a reliable estimator should produce overlapping point estimates for $S$—with
varying variance but consistent expectation when evaluated over equal-length
campaigns.

We propose to evaluate whether estimators of maximum reachability $S$ yield
consistent point estimates with overlapping confidence intervals, while allowing
for differences in estimation accuracy (i.e., variability) across fuzzing
campaigns with different sampling unit definitions $r$.


\section{Execution Plan}
\label{sec:setup}
We describe the details our proposed experiment here.
\subsection{Research Questions}
How well do the current techniques for estimating the maximum reachability
of statements in a program perform? What are their limitations? Can the
point estimators be trusted? and are the confidence intervals obtained accurate?
The following research question will be investigating each of these points.

\noindent\textbf{RQ1:} How accurate are the estimators of maximum reachability
in terms of the point estimate as well as the CIs?

The second question that we want to be answered is regarding the reliability
of the sampling units used in species richness based statistical estimators for
maximum reachability. We note that there are several methods for formulating
the problem of maximum reachability as a species richness estimate. For example,
if one is using traditional test cases, the sampling unit can be the test
classes and test methods as Kuznetsov et al.~\cite{Kuznetsov2024empirical}
demonstrates. On the other hand, in the case of fuzzing, where there is no
simple mechanism to identify non-overlapping sampling units, one can use the
incidence data under various time intervals as
Liyanage et al.~\cite{liyanage2023reachable} demonstrates. In both these cases, the
sampling units can be varied. In particular, Liyanage et al. chose 15~minutes
as the time interval for sampling unit. However, it can be asked if a time
interval of 5 minutes or even 60 minutes would have similar estimates.
The following research question will investigate this question.

\noindent\textbf{RQ2:} Are the maximum reachability estimators for $S$
reliable when sampling unit size $r$ is changed?


\subsection{Estimators of Reachable Coverage}

We leverage 12 state-of-the-art biostatistical species richness estimators, as employed in Liyanage et al.~\cite{liyanage2023reachable} and Kuznetsov et al.~\cite{Kuznetsov2024empirical}. Based on their underlying construction, several estimators can be grouped into distinct classes. Table~\ref{tab:estimators} summarizes all estimators along with the corresponding notations used throughout this report.

\begin{table}[ht]
\centering
\caption{List of Reachability Estimators}
\label{tab:estimators}
\scriptsize
\begin{tabular}{@{}p{3cm}p{4cm}p{1.2cm}@{}}
\toprule
\textbf{Estimator class} & \textbf{Estimator} & \textbf{Notation} \\
\midrule
\multirow{3}{*}{\shortstack[l]{Chao-type}} 
    & Chao2~\cite{chao1987estimating} & $\hat{S}_{\text{Chao2}}$ \\
    & Bias corrected Chao2 (Chao2\_bc)~\cite{chao2005new} & $\hat{S}_{\text{Chao2\_bc}}$ \\
    & Improved Chao2 (iChao2)~\cite{chiu2014improved} & $\hat{S}_{\text{iChao2}}$ \\
\midrule
\multirow{2}{*}{\shortstack[l]{Jackknife}} 
    & First-order Jackknife (JK1)~\cite{burnham1978estimation} & $\hat{S}_{\text{JK1}}$ \\
    & Second-order Jackknife (JK2)~\cite{burnham1978estimation} & $\hat{S}_{\text{JK2}}$ \\
\midrule
\multirow{2}{*}{\shortstack[l]{Incidence-based\\coverage estimators (ICE)}} 
    & ICE~\cite{lee1994estimating} & $\hat{S}_{\text{ICE}}$ \\
    & ICE1~\cite{gotelli2013measuring} & $\hat{S}_{\text{ICE1}}$ \\
\midrule
\multirow{1}{*}{\shortstack[l]{Zelterman estimator}} 
    & Zelterman~\cite{bohning2010some} & $\hat{S}_{\text{Zelterman}}$ \\
\midrule
\multirow{1}{*}{\shortstack[l]{Bootstrap estimator}} 
    & Bootstrap~\cite{smith1984nonparametric} & $\hat{S}_{\text{Bootstrap}}$ \\
\midrule
\multirow{1}{*}{\shortstack[l]{Frequency-based}} 
    & Chao-Bunge~\cite{chao2002estimating} & $\hat{S}_{\text{Chao-Bunge}}$ \\
\midrule
\multirow{1}{*}{\shortstack[l]{Unconditional nonparametric \\ maximum likelihood estimator}} 
    & UNPMLE~\cite{norris1998non} & $\hat{S}_{\text{UNPMLE}}$ \\
    & & \\
\midrule
\multirow{1}{*}{\shortstack[l]{Penalized nonparametric \\ maximum likelihood estimator}} 
    & PNPMLE~\cite{wang2005penalized} & $\hat{S}_{\text{PNPMLE}}$ \\
    & & \\
\bottomrule
\end{tabular}
\end{table}

For the exact formulations of these estimators, we refer to~\cite{liyanage2023reachable} and~\cite{Kuznetsov2024empirical}, where these estimators were originally applied in the context of software testing.

\subsection{Dataset: Fuzzer, Subject Programs, Seed Corpus}

\noindent\textbf{Fuzzer.} For our experiments,
we employ AFL++~\cite{fioraldi2020AFL++}, a widely adopted
state-of-the-art greybox fuzzer known for its effectiveness and prevalent use in
recent reachability estimation studies.
Unlike its predecessor AFL~\cite{zalewski2017american}, which selects seeds from a circular queue, AFL++ implements a probabilistic seed selection mechanism that aligns with the probabilistic model in STADS described in \Cref{sec:preliminaries}.

\noindent\textbf{Programs.} As detailed in Section~\ref{sec:method}, we will generate a
diverse set of large C parsers with known reachability as our
fuzzing benchmarks. For this research, we will generate 100
parsers, with an average 100,000 lines of code in each.
This is similar to the size of subjects in
Fuzzbench~\cite{metzman2021fuzzbench}.
We will ensure that the parsers have complex control-flows with
loops, conditionals, direct, indirect, and linear recursion.
We will also ensure that these targets are compatible with AFL++ and suitable for
large-scale fuzzing experiments.

\noindent\textbf{Campaign length and trials.}
We will fuzz each target at least 24 hours per previous protocols~\cite{manes2019the}.
For each target, we conduct multiple independent fuzzing trials $K=100$ to ensure consistency and reduce the impact of random variation.

\noindent\textbf{Initial seed corpus.} As the introduced benchmark programs lack well-established seed corpora, we use a well-formed valid inputs generated from the respective grammars
as the initial seed corpus for each subject over multiple fuzzing trials.

\subsection{Evaluation Metrics}
To assess the performance of statistical estimators of maximum reachability, we adopt
the following metrics for \textbf{RQ1}. 

\noindent\textbf{Mean bias.} We define the bias at arbitrary sampling point $t$ as the difference between the estimator's
point estimate $\hat{S}$ and the known ground truth $S$ available from the synthetic benchmark. We compute the average of relative bias over $K$ trials to account for random error.
Formally,
\begin{equation}
\text{\emph{mean bias}}(t) = \sum_{i=1}^K \frac{\hat{S}_i(t)-S}{KS}
\end{equation}
A smaller absolute bias indicates better estimate accuracy. 

\noindent\textbf{Variance across multiple runs.} Variance quantifies the stability of the
estimator across multiple fuzzing campaigns. For each estimator, we will report the \emph{variance} or \emph{imprecision}\footnote{Imprecision~\cite{liyanage2023reachable} is defined as the variance of individual bias values.}
of $\hat{S}$ over repeated $K$ trials. Low variance/imprecision indicates greater stability and reliability of the estimation.

\noindent\textbf{Confidence interval (CI) coverage.} When estimators provide confidence intervals, we will evaluate their coverage by checking whether the true $S$ lies within the specified interval. We will report the proportion of CIs containing the true $S$.

To address \textbf{RQ2}, we adopt
\textbf{Sampling unit sensitivity.} We will consider an estimator reliable if its
point estimates across different time intervals have overlapping confidence intervals, and no significant shifts in mean estimates. We will use Welch's t-test to determine significant differences. A reliable estimator should result in no significant
differences. We will verify normality of data, and fall back to Mann-Whitney U-test if significant non-normality is observed.


\section{Related Work}
\label{sec:related}

Determining reachability is of fundamental importance in software testing~\cite{liyanage2023reachable}.
However, establishing the reachability of certain code regions is particularly challenging in large,
code bases~\cite{latoza2010developers}.
Static analysis methods, such as symbolic execution, often suffer from over- or under-approximation when estimating reachable coverage~\cite{liyanage2023reachable,aniche2015why}.
For instance, Nikoli\'{c} and Spoto~\cite{nikolic2013reachability} proposed approximating the reachability of program variables as a new abstract domain for static analysis. While their approach yields over-approximations, authors argue that it can be conservatively applied to identify unreachable code. Similarly, Mikol\'{a}\v{s} et al.~\cite{janota2007reachability} leveraged annotated code to define unreachability conditions and proposed an efficient algorithm for detecting unreachable code.

To assess reachability in dynamic analysis, both constraint-solving techniques such as SMT
and data-driven estimation methods have been explored. Naus et al.~\cite{naus2023low} proposed a technique for automatically generating preconditions to trigger specific post-conditions (e.g., bugs) using low-level code analysis. Similarly, Liew et al.~\cite{liew2019just} demonstrated how to encode SMT formulas within coverage-guided fuzzers to discover inputs that reach targeted program locations. In contrast, statistical approaches directly attempt to estimate the test effectiveness (aka maximum reachability) through observed behaviors. The pioneering STADS framework~\cite{boehme2018stads} introduced a suite of bio-statistical estimators by modeling fuzzing as a statistical sampling process. A recent evaluation of these estimators highlighted key challenges associated with reachability estimation~\cite{liyanage2023reachable}. When applied to estimate the number of killable mutants in mutation analysis, Kuznetsov et al.~\cite{Kuznetsov2024empirical} empirically demonstrated the unreliability of STADS estimators due to variations in sampling unit definitions. This finding, together with other recent studies, underscores the need for \emph{structure-aware} estimators that incorporate internal program structure, as opposed to structure-agnostic alternatives~\cite{lee2023statistical}.

Using synthetic benchmarks by program generation has been attempted before.
Fuzzle~\cite{lee2022fuzzle} is a benchmark for fuzzer evaluation that is built
by encoding a sequence of moves in a maze as a chain of function calls. The
limitation here is that mazes are rarely similar to real-world programs.
They are limited by the moves one can make in a maze. Furthermore, solving
a maze typically requires finding a single path through a maze.
Recursive descent parsers, on the other hand, are real-world programs, and
are one of the major subjects in practical fuzzing.
Olympia~\cite{chadt2024olympia} based on Fuzzle generates \emph{solidity contracts} from generated mazes.

\section{Future Work}\label{sec:future}
In this work, we proposed evaluation of reachability estimators
using generated programs as subjects.
Recent research has indicated that while estimation of maximum reachability
may be inaccurate in the beginning, one can reliabily estimate whether it is
now time to stop fuzzing based on when the doubleton count reaches or surpasses
singleton count~\cite{liyanage2023reachable}. Verifying this criteria would be
the next step in our research.

\section{Threats to Validity}\label{sec:threats}

As with any empirical study, our findings on reachable coverage estimation in fuzzing are subject to threats to validity.

\emph{\textbf{Internal validity}} concerns the impact of systematic errors.
Implementation or tooling defects could introduce bias in our findings.
To mitigate, we carefully validate our tool implementations,
data collection, and processing pipelines.

To reduce the impact of random variance from fuzzer choice,
seed selection, and grammar parameters, we perform multiple independent
runs and report aggregated results.

Finally, the number of inputs possible in a given time interval
can be impacted by the load on the machine on which the experiment is run. We
will mitigate this impact by carefully controlling the machine load during the
experiments.


\emph{\textbf{External validity}} refers to generalizability of our findings.
Using synthetic parser benchmarks limits the generalization of our findings
to parser-like programs in terms of the complexity of control-flow. We
acknowledge this threat, and point out that parsers themselves are a subject of
fuzzing in the real-world, and are considered standard programs in many benchmarks.
\rebuttal{We use AFL++ which limits the programs being fuzzed to C/C++. While this can potentially impact the generalizability, we argue that the control flow structures we synthesize are common across a variety of languages.}

Finally, the second part of our study proposes to use real-world programs, which
can mitigate the impact of the synthetic benchmarks in the generality of our findings. 


\emph{\textbf{Construct validity}} refers to the extent to which the measurement
actually captures the intent. We note that in our first benchmark, we are
\emph{directly} measuring the reachability of program elements, and the
ground-truth is known. Hence, we are not impacted by a threat to validity in the
first experiment. However, the second experiment is indirect, and relies on the
assumption that sampling units should not have a significant impact on the
point estimate. We acknowledge this threat, and note that this assumption is
consistent with prior work \cite{Kuznetsov2024empirical} using similar methodologies.


\section{Conclusion}
\label{sec:conclusion}
The budget allocated to fuzzing, and the campaign stopping criteria is informed
by the maximum achievable coverage. However, statically inferring
reachability in real-world programs is impossible. Hence, statistical estimators
are often relied upon.
In this report, we propose to assess the reliability of species estimators when
they are used for estimating reachable coverage. We propose to do that both by
providing a synthetic benchmark with labeled ground truth, and also by
using a separate means of checking the reliability of estimators by varying
the sampling units used.

\rebuttal{
\section{Acknowledgments}
This research was funded by Australian Research Council Discovery Project DP210101984.
}

\balance
\bibliographystyle{IEEEtran}
\bibliography{icsme-registered.bib}

\begin{thebibliography}{10}
\providecommand{\url}[1]{#1}
\csname url@samestyle\endcsname
\providecommand{\newblock}{\relax}
\providecommand{\bibinfo}[2]{#2}
\providecommand{\BIBentrySTDinterwordspacing}{\spaceskip=0pt\relax}
\providecommand{\BIBentryALTinterwordstretchfactor}{4}
\providecommand{\BIBentryALTinterwordspacing}{\spaceskip=\fontdimen2\font plus
\BIBentryALTinterwordstretchfactor\fontdimen3\font minus
  \fontdimen4\font\relax}
\providecommand{\BIBforeignlanguage}[2]{{%
\expandafter\ifx\csname l@#1\endcsname\relax
\typeout{** WARNING: IEEEtran.bst: No hyphenation pattern has been}%
\typeout{** loaded for the language `#1'. Using the pattern for}%
\typeout{** the default language instead.}%
\else
\language=\csname l@#1\endcsname
\fi
#2}}
\providecommand{\BIBdecl}{\relax}
\BIBdecl

\bibitem{boehme2016coverage}
M.~B\"{o}hme, V.-T. Pham, and A.~Roychoudhury, ``Coverage-based greybox fuzzing
  as markov chain,'' ser. CCS '16, 2016, p. 1032–1043.

\bibitem{fell2017review}
J.~Fell, ``A review of fuzzing tools and methods,'' \emph{PenTest Magazine},
  2017.

\bibitem{fraser2013whole}
G.~Fraser and A.~Arcuri, ``Whole test suite generation,'' \emph{IEEE
  Transactions on Software Engineering}, vol.~39, no.~2, pp. 276--291, 2013.

\bibitem{horgan1994achieving}
J.~R. Horgan, S.~London, and M.~R. Lyu, ``Achieving software quality with
  testing coverage measures,'' \emph{Computer}, vol.~27, no.~9, 1994.

\bibitem{boehme2018stads}
M.~B\"{o}hme, ``{STADS}: Software testing as species discovery,'' \emph{ACM
  TOSEM}, vol.~27, no.~2, pp. 7:1--7:52, Jun. 2018.

\bibitem{chao2016species}
A.~Chao, C.-H. Chiu \emph{et~al.}, ``Species richness: estimation and
  comparison,'' \emph{Wiley StatsRef: statistics reference online}, vol.~1,
  p.~26, 2016.

\bibitem{liyanage2021security}
D.~Liyanage, ``Security guarantees for automated software testing,'' in
  \emph{ESEC/FSE}, ser. ESEC/FSE, 2021.

\bibitem{liyanage2023reachable}
D.~Liyanage, M.~B\"{o}hme, C.~Tantithamthavorn, and S.~Lipp, ``Reachable
  coverage: Estimating saturation in fuzzing,'' in \emph{ICSE}, ser. ICSE,
  2023, p. 371–383.

\bibitem{Kuznetsov2024empirical}
K.~Kuznetsov, A.~Gambi, S.~Dhiddi, J.~Hess, and R.~Gopinath, ``Empirical
  evaluation of frequency based statistical models for estimating killable
  mutants,'' in \emph{ESEM}, ser. ESEM, 2024, p. 61–71.

\bibitem{boehme2021residual}
M.~B\"{o}hme, D.~Liyanage, and V.~W\"{u}stholz, ``Estimating residual risk in
  greybox fuzzing,'' in \emph{ESEC/FSE}, ser. ESEC/FSE, 2021.

\bibitem{chao2017thirty}
A.~Chao and R.~K. Colwell, ``Thirty years of progeny from chao's inequality:
  Estimating and comparing richness with incidence data and incomplete
  sampling,'' \emph{SORT: statistics and operations research transactions},
  vol.~41, no.~1, pp. 0003--54, 2017.

\bibitem{nguyen2022bedivfuzz}
H.~L. Nguyen and L.~Grunske, ``Bedivfuzz: Integrating behavioral diversity into
  generator-based fuzzing,'' in \emph{ICSE}, ser. ICSE, 2022, pp. 1--13.

\bibitem{boehme2020boosting}
M.~B{\"o}hme, V.~Man{\`e}s, and S.~K. Cha, ``Boosting fuzzer efficiency: An
  information theoretic perspective,'' in \emph{ESEC/FSE}, ser. ESEC/FSE, 2020.

\bibitem{colwell2012models}
R.~K. Colwell, A.~Chao, N.~J. Gotelli, S.-Y. Lin, C.~X. Mao, R.~L. Chazdon, and
  J.~T. Longino, ``Models and estimators linking individual-based and
  sample-based rarefaction, extrapolation and comparison of assemblages,''
  \emph{Journal of plant ecology}, vol.~5, no.~1, pp. 3--21, 2012.

\bibitem{chao1987estimating}
A.~Chao, ``Estimating the population size for capture-recapture data with
  unequal catchability,'' \emph{Biometrics}, pp. 783--791, 1987.

\bibitem{chao2005new}
A.~Chao, R.~L. Chazdon, R.~K. Colwell, and T.-J. Shen, ``A new statistical
  approach for assessing similarity of species composition with incidence and
  abundance data,'' \emph{Ecology letters}, vol.~8, no.~2, pp. 148--159, 2005.

\bibitem{chiu2014improved}
C.-H. Chiu, Y.-T. Wang, B.~A. Walther, and A.~Chao, ``An improved nonparametric
  lower bound of species richness via a modified good--turing frequency
  formula,'' \emph{Biometrics}, vol.~70, no.~3, pp. 671--682, 2014.

\bibitem{burnham1978estimation}
K.~P. Burnham and W.~S. Overton, ``Estimation of the size of a closed
  population when capture probabilities vary among animals,''
  \emph{Biometrika}, pp. 625--633, 1978.

\bibitem{lee1994estimating}
S.-M. Lee and A.~Chao, ``Estimating population size via sample coverage for
  closed capture-recapture models,'' \emph{Biometrics}, pp. 88--97, 1994.

\bibitem{gotelli2013measuring}
N.~J. Gotelli and A.~Chao, ``Measuring and estimating species richness, species
  diversity, and biotic similarity from sampling data,'' \emph{Encyclopedia of
  biodiversity}, pp. 195--211, 2013.

\bibitem{bohning2010some}
D.~B{\"o}hning, ``Some general comparative points on chao's and zelterman's
  estimators of the population size,'' \emph{Scandinavian Journal of
  Statistics}, vol.~37, no.~2, pp. 221--236, 2010.

\bibitem{smith1984nonparametric}
E.~P. Smith and G.~van Belle, ``Nonparametric estimation of species richness,''
  \emph{Biometrics}, pp. 119--129, 1984.

\bibitem{chao2002estimating}
A.~Chao and J.~Bunge, ``Estimating the number of species in a stochastic
  abundance model,'' \emph{Biometrics}, vol.~58, no.~3, pp. 531--539, 2002.

\bibitem{norris1998non}
J.~L. Norris and K.~H. Pollock, ``Non-parametric mle for poisson species
  abundance models allowing for heterogeneity between species,''
  \emph{Environmental and Ecological Statistics}, vol.~5, pp. 391--402, 1998.

\bibitem{wang2005penalized}
J.-P.~Z. Wang and B.~G. Lindsay, ``A penalized nonparametric maximum likelihood
  approach to species richness estimation,'' \emph{Journal of the American
  Statistical Association}, vol. 100, no. 471, pp. 942--959, 2005.

\bibitem{fioraldi2020AFL++}
A.~Fioraldi, D.~Maier, H.~Ei{\ss}feldt, and M.~Heuse, ``{AFL++: Combining
  Incremental Steps of Fuzzing Research},'' in \emph{USENIX Workshop on
  Offensive Technologies}, 2020.

\bibitem{zalewski2017american}
\BIBentryALTinterwordspacing
M.~Zalewski, ``American fuzzy lop,'' 2017. [Online]. Available:
  \url{https://github.com/google/AFL}
\BIBentrySTDinterwordspacing

\bibitem{metzman2021fuzzbench}
J.~Metzman, L.~Szekeres, L.~Simon, R.~Sprabery, and A.~Arya, ``Fuzzbench: an
  open fuzzer benchmarking platform and service,'' in \emph{ESEC/FSE}, 2021,
  pp. 1393--1403.

\bibitem{manes2019the}
V.~J. Man{\`e}s, H.~Han, C.~Han, S.~K. Cha, M.~Egele, E.~J. Schwartz, and
  M.~Woo, ``The art, science, and engineering of fuzzing: A survey,''
  \emph{IEEE TSE}, vol.~47, no.~11, pp. 2312--2331, 2019.

\bibitem{latoza2010developers}
T.~D. LaToza and B.~A. Myers, ``Developers ask reachability questions,'' in
  \emph{ICSE}, ser. ICSE, 2010, p. 185–194.

\bibitem{aniche2015why}
M.~F. Aniche, G.~A. Oliva, and M.~A. Gerosa, ``Why statically estimate code
  coverage is so hard? a report of lessons learned,'' in \emph{2015 29th
  Brazilian Symposium on Software Engineering}, 2015, pp. 185--190.

\bibitem{nikolic2013reachability}
u.~Nikoli\'{c} and F.~Spoto, ``Reachability analysis of program variables,''
  \emph{ACM TPLS}, vol.~35, no.~4, Jan. 2014.

\bibitem{janota2007reachability}
M.~Janota, R.~Grigore, and M.~Moskal, ``Reachability analysis for annotated
  code,'' in \emph{ESEC/FSE}, ser. SAVCBS '07.\hskip 1em plus 0.5em minus
  0.4em\relax New York, NY, USA: Association for Computing Machinery, 2007, p.
  23–30.

\bibitem{naus2023low}
N.~Naus, F.~Verbeek, M.~Schoolderman, and B.~Ravindran, ``Low-level
  reachability analysis based on formal logic,'' in \emph{International
  Conference on Tests and Proofs}.\hskip 1em plus 0.5em minus 0.4em\relax
  Springer, 2023, pp. 21--39.

\bibitem{liew2019just}
D.~Liew, C.~Cadar, A.~F. Donaldson, and J.~R. Stinnett, ``Just fuzz it: solving
  floating-point constraints using coverage-guided fuzzing,'' in
  \emph{ESEC/FSE}, ser. ESEC/FSE 2019, 2019, p. 521–532.

\bibitem{lee2023statistical}
\BIBentryALTinterwordspacing
S.~Lee and M.~B\"{o}hme, ``Statistical reachability analysis,'' in
  \emph{ESEC/FSE}, ser. ESEC/FSE.\hskip 1em plus 0.5em minus 0.4em\relax New
  York, NY, USA: Association for Computing Machinery, 2023. [Online].
  Available: \url{https://doi.org/10.1145/3611643.3616268}
\BIBentrySTDinterwordspacing

\bibitem{lee2022fuzzle}
H.~Lee, S.~Kim, and S.~K. Cha, ``Fuzzle: Making a puzzle for fuzzers,'' in
  \emph{Proceedings of the 37th IEEE/ACM International Conference on Automated
  Software Engineering}, 2022, pp. 1--12.

\bibitem{chadt2024olympia}
\BIBentryALTinterwordspacing
J.~Chadt, C.~Hochrainer, V.~W\"{u}stholz, and M.~Christakis, ``Olympia: Fuzzer
  benchmarking for solidity,'' ser. ASE '24.\hskip 1em plus 0.5em minus
  0.4em\relax New York, NY, USA: Association for Computing Machinery, 2024, p.
  2362–2365. [Online]. Available:
  \url{https://doi.org/10.1145/3691620.3695352}
\BIBentrySTDinterwordspacing

\end{thebibliography}

\end{document}